\documentclass[%
 twocolumn,
superscriptaddress,
groupedaddress,
showpacs,preprintnumbers,
nofootinbib,
 amsmath,amssymb,
 aps,
 prd,
 letterpaper,
 tightenlines,
showkeys,
floats,
floatfix,
]{revtex4-1}

\usepackage{graphicx}
\usepackage{dcolumn}
\usepackage{bm}
\usepackage{color}

\usepackage{color}

\begin{document}

\preprint{}

\title{Cosmological constraints in a two branes system for a vacuum bulk}

\author{Juan Luis P\'erez} 
\email{einstein1\_25@fisica.ugto.mx}
\affiliation{Departamento de F\'isica, DCI, Campus Le\'on, Universidad de Guanajuato, 37150, Le\'on, Guanajuato, M\'exico}
\author{Luis A. Ure\~na L\'opez}
\email{lurena@fisica.ugto.mx}
\affiliation{Departamento de F\'isica, DCI, Campus Le\'on, Universidad de Guanajuato, 37150, Le\'on, Guanajuato, M\'exico}
\author{Rub\'en Cordero}
\email{rcordero@esfm.ipn.mx}
\affiliation{Departamento de F\'isica, Escuela Superior de F\'isica y Matem\'aticas del IPN, Unidad Adolfo L\'opez Mateos, Edificio 9, 07738, M\'exico, Distrito Federal, M\'exico}

\date{\today}

\begin{abstract}
We present the study of a two 3-branes system embedded in a 5-dimensional space-time in which the fifth dimension is compacted on a $S^{1}/Z_{2}$ orbifold and the space between the two branes is vacuum. Assuming isotropic and homogeneous space time, we probe that the dynamics of the visible brane is affected by the dynamics of the hidden brane and the possible components which fills it. We study two different cases: assuming a constant equation of state and inflationary scalar field, and its possible consequences in the brane evolution.
\end{abstract}


\maketitle

\section{Introduction}

Brane-worlds models it is an interesting viewpoint of the universe dynamics adding new degrees of freedom which can help to solve the problems of dark matter (DM) and dark energy (DE). In principle, these models has been motivated by String theory and M-theory \cite{Schwarz:2008kd,cuerdas1,cuerdas2,cuerdas3}, where our visible universe can be seen as a 4D manifold (brane) immersed in a space-time of more than three spatial dimensions (bulk)  where usually the standard model of particles (SM) fields are trapped on the brane, being the gravity the only field that escape to the bulk.

Historically, the models with more spatial dimensions begin with the Kaluza and Klein works \cite{k-k-1,k-k-2}, who's, following the Nordstrom idea, built a 5D theory as an attempt for unify gravity and electromagnetic forces. A novel feature of 5D models is that $M_{p}=M_{4}$, the 4D Planck scale, is not more the fundamental scale, which is $M_{5}$; additionally, the compact extra dimensions implies every multi-dimensional field corresponds to a KaluzaﾐKlein tower of four-dimensional particles with increasing masses. At low energies, only massless (at $E >> 1/R$) particles can be produced, whereas at $E \approx1/R$ extra dimensions are detectable. This is the starting point of string theories which, trying to reconcile quantum mechanics and general relativity, postulate a more general space-time of 4+D dimensions, where the fundamental particles are conceived as small vibrating strings. 

Recent research in string theory and its generalization M-theory have suggested the number of dimensions, to make a consistent quantum  string theory is eleven. Inherited in these models are the p-branes ($0<p<9$), which are the fundamental constituents of the universe. A brane, where the open strings have their endpoints, is called D-brane. Our visible universe can be a very large D-brane extending over three spatial dimensions. Material objects, made of open strings, are confined on such D-brane, while gravity and other exotic matter such as the dilaton can propagate in the bulk. This scenario is called brane cosmology or brane-world cosmology A reduction to 5D of M-theory is suggested by Horava and Witten \cite{HW}. 

The strong coupling limit of the $E_{8}\times E_{8}$ heterotic string theory at low energy is described by 11D supergravity with the eleventh dimension compactified on an $S_{1}/Z_{2}$ orbifold. The two boundaries of space-time are two 10ﾐbranes, on which gauge theories are confined. Witten argued that 6 of the 11 dimensions can be consistently compactified on a CalabiﾐYau threefold and that the size of the Calabi-Yau manifold
can be substantially smaller than the space between the two boundary branes. Thus,
in that limit spaceﾐtime looks fiveﾐdimensonal. A 5D realization of the HW theory and the corresponding brane-world cosmology is given in \cite{300,301,302}. These solutions can be thought of as effectively 5D, with an extra dimension that
can be large relative to the fundamental scale, providing the basis
for the Arkani-Dimopoulos-Dvali (ADD) \cite{ADD-1}, Randall-Sundrum (RS) \cite{randall,randall2}, and Dvali-Gabadadze-Porrati (DGP) brane models of 5D gravity.

Originally, the brane-worlds models attempted to solve the hierarchy
problem, namely the large difference in magnitudes
between the Planck and electroweak scales, $M_{pl}/M_{EW}\simeq10^{16}$.Using the large extra dimensions, Arkani \emph{et.al.} \cite{ADD-1} suppose a D-dimensional bulk with
the Planck scale $M_{D}$. The 4-dimensional Planck
mass $M_{pl}$  is given by $M^{2}_{pl}=V_{D-4}M^{D-2}_{D}$ , where $V_{D-4}$
is the volume of the $(D-4)$-dimensional space.
If the extra dimensions are large enough, even
$M_{D}$ is in the order of electroweak scale $M_{D}\simeq M_{EW}\simeq TeV$ , one can get the correct order of $M_{pl}\simeq10^{16} TeV$ ,
whereby the hierarchy problem is resolved. In the Randall Sundrum I (RSI)
model, the mechanism is completely different \cite{randall}. Instead
of using large dimensions, RS used the warped factor
$\sigma(y) = k\left|y\right|$, for which the mass $m_{0}$ measured on
the invisible (Planck) brane is related to the mass $m$
measured on the visible (TeV) brane by $m = e^{-ky_{c}}m_{0}$.
Clearly, by properly choosing the distance $y_{c}$ between the
two branes, one can lower $m$ to the order of TeV , even
$m_{0}$ is still in the order of $M_{pl}$. It should be noted that
the five-dimensional Planck mass $M_{5}$ in the RS1 scenario
is still of the order of $M_{pl}$ and the two are related by
$M^{2}_{pl} = M^{3}k^{-1}\left(1-e^{-2ky_{c}}\right)\simeq M^{2}_{5}$ for $k\simeq M_{5}$. 

In the context of two-brane models with matter, one can naturally to ask if the parameters which determines the evolution in time in both branes are related. Binetruy \emph{et al.} \cite{Binetruy:1999ut} has showed that there exists an equation which relate the fields in both branes assuming a mutual interaction between them, through a topological constrains. For example, in \cite{aspeita1,aspeita2} the authors assume that the hidden brane is dominated by a scalar field, trying to reproduce the dark matter effect in the visible brane.

Based in the RS models and in the previous results founded by \cite{Langlois,perez1,perez2}, this paper focus in generalize the solution founded in \cite{Langlois} for a vacuum 5D bulk in which we propose the metric coefficients with a particular mathematical structure. This formalism generates a dynamical equation for the Hubble parameter in our brane $H_{c}$ closely related with the dynamics of the hidden brane $H_{0}$; in the other words, the fields immersed in the hidden brane generates dynamics in the visible brane through gravitational effects. First, we study a toy model with equations of state (EoS) constant and the repercussions in the mutual brane evolution; on the other hand we study as a particular field election a scalar field as responsible of the inflationary dynamics in the hidden brane.  

The present paper is organized as follows: In Sec. \ref{Sec-2} we explore about the mathematical solutions for a system with two branes with a cosmological constant between them, obtaining from the 5-dimensional Einstein equations, the corresponding differential equations and boundary conditions for the two branes. We solve the aforementioned equations and found a family of exact solutions for the metric coefficients.
In Sec. \ref{Sec-3} we apply the solutions to the boundary conditions and give the equations that relates the cosmology in both branes; a particular cases are studied for a single component, inflationary scalar field in the hidden brane, and the effect on the visible brane. Finally, the conclusions and remarks are presented in \ref{concl}. 

We will henceforth use units in which $c=\hbar=1$.

\section{Two branes embedded in a 5D Spacetime.} \label{Sec-2}

We propose the basic equations for this model assuming two branes (visible and hidden) embedded in a five dimensional manifold. We start writing the five dimensional action as

\begin{equation}
S[g_{(5)}]=-\frac{1}{2\kappa^{2}_{(5)}}\int d^{5}x\sqrt{-g_{(5)}}R_{(5)}\pm\sum_{i}\int d^{5}x\sqrt{-g_{(5)}}\mathcal{L}_{i}
\end{equation}
where $g_{(5)}$ is the five dimensional metric, $\kappa_{(5)}$ is the five dimensional gravitational constant, $R_{(5)}$ is the five dimensional Ricci scalar and $\mathcal{L}$ corresponds to the scalar field Lagrangian for the visible and hidden brane respectively. We propose that the two branes are located in $y=0$ and $y=y_{c}$ respectively, both brane are immersed in a homogeneous and isotropic 5-dimensional manifold where the most general metric can be described as

\begin{equation}
ds^{2} =-n^{2}(t,\left|y\right|) dt^{2} + a^{2}(t,\left|y\right|) g_{i j} dx^{i} dx^{j} +b^{2}(t,\left|y\right|) dy^{2}. \label{metrica}
\end{equation}
As an important feature, we impose the symmetries enumerated in the following way:

\begin{enumerate}

\item Reflection, $(x^{\mu},y) \rightarrow (x^{\mu},-y)$

\item Compactification, $(x^{\mu},y)\to(x^{\mu},y+2iy_{c}), \, i=1,2,\ldots$

\end{enumerate}
Similarly, we demand that each metric coefficients $a(t,\left|y\right|)$, $n(t,\left|y\right|)$ and $b(t,\left|y\right|)$ are subjected to the conditions \cite{Wang}

\begin{equation}
\label{condicion1}	
\left[F^{\prime}\right]_{0}=2F^\prime |_{y=0+},
\end{equation}

\begin{equation}
\label{condicion2}	
\left[F^{\prime}\right]_{c}=-2F^\prime |_{y=y_{c}-},
\end{equation}

\begin{equation}
\label{condicion3} F^{\prime \prime}=\frac{d^{2}F(t,\left| y \right|)}{d\left| y \right|^{2}} + \left[ F^\prime\right]_{0} \delta(y) + \left[ F^\prime \right]_{c} \delta(y-y_{c}),
\end{equation}
where the prime denotes derivate with respect to $y$, the square brackets denotes the discontinuity in the first derivative at the positions $y=0$ and $y=y_{c}$ and $F$ is a generic function which meets the above conditions \cite{Wang}. 

The equation \eqref{condicion3} is obtained if we demand that $d\left|y\right|/dy=1$, and $d^{2} \left| y \right|/dy^{2} = 2\delta(y) -2\delta(y-y_{c})$, for $y\in[0,y_{c}]$. The subindex $0$ will be used for quantities valued at $y=0$, whereas a subindex $c$ will be used for quantities valued at $y=y_{c}$. Now, in order to obtain exact dynamical solutions, we write the five-dimensional Einstein equations, (See Appendix \ref{Ap-1}), for the
metric \eqref{metrica} together with the energy momentum tensor as

\begin{eqnarray}
\label{tme} \tilde{T}^{^{A}}_{_{B}}&=&\hat{T}^{^{A}}_{_{B}}+\frac{\delta(y)}{b_{0}}diag(-\rho_{0},\textbf{p}_{0},0)\nonumber\\&&+\frac{\delta(y-y_{0})}{b_{c}}diag(-\rho_{c},\textbf{p}_{c},0),
\end{eqnarray}
where the first term corresponds to the bulk contribution and the second and third term corresponds to the branes embedded in the 5D manifold. Usually the term $\hat{T}^{^{A}}_{_{B}}$ is in the form of a five dimensional cosmological constant, namely

\begin{equation}\label{Lambda-5}
\hat{T}_{_{AB}}=-\frac{\Lambda_{5}}{\kappa^{2}_{_{(5)}}}g_{_{AB}},
\end{equation}
where using the equation $\nabla_{_{A}}\tilde{G}^{^{A}}_{_{B}}=0$, immediately yields the conservation equation 

\begin{equation}
\label{conservacion}
\dot{\rho}_{0}+3(p_{0}+\rho_{0})\frac{\dot{a}_{0}}{a_{0}}=0.
\end{equation}
According with the Israel \textit{junction conditions} \cite{PhysRevD.43.1129}, we describe the presence of an energy density in terms of a discontinuity in the metric across the origin in the extra coordinate. So,
following Wang notation \cite{Wang} we obtain the metric coefficients that satisfy the following boundary conditions

\begin{eqnarray}
\label{salto1}
\frac{\left[a'\right]_{0}}{a_{0}b_{0}}&=&-\frac{\kappa^{2}_{(5)}}{3}\rho_{0}, \\
\label{salto2}
\frac{\left[n'\right]_{0}}{n_{0}b_{0}}&=&\frac{\kappa^{2}_{(5)}}{3}(3p_{0}+2\rho_{0}).
\end{eqnarray}
This implies that the \textit{jump} in the first derivate of the metric coefficients is proportional to the energy density across the origin. Similarly, in $y=y_{c}$, we have

\begin{eqnarray}
\label{salto3}
\frac{\left[a'\right]_{c}}{a_{c}b_{c}}&=&-\frac{\kappa^{2}_{(5)}}{3}\rho_{c}, \\
\label{salto4}
\frac{\left[n'\right]_{c}}{n_{c}b_{c}}&=&\frac{\kappa^{2}_{(5)}}{3}(3p_{c}+2\rho_{c}).
\end{eqnarray}
Following Langlois \cite{Langlois}, the $(0,0)$ component of Einstein equation (See Appendix \ref{Ap-1}) gives
\begin{equation}\label{friedmann3}
\left(\frac{a'}{ab}\right)^2-\left(\frac{\dot{a}}{an}\right)^2 =ka^{-2}-\frac{\Lambda_{5}}{6}+C_{_{DR}}a^{-4}, 
\end{equation}
where $C_{_{DR}}$ is the called \emph{dark radiation} contribution \cite{Maartens:2010ar}. On the other hand, $\tilde{G}_{05}=\kappa^{2}_{(5)}\tilde{T}_{05}=0$ (which physically means that there is no flow of matter along the fifth dimension) implies

\begin{equation}\label{0-5}
\frac{\dot{b}}{b}=\frac{n}{a'}\left[\frac{\dot{a}}{n}\right]'.
\end{equation}
Using the previous equations, in the following subsections of the paper we solve the equations (\ref{friedmann3}) and (\ref{0-5}) in different scenarios.

To obtain the exact solutions of the equations (\ref{friedmann3}) and (\ref{0-5}) in a vacuum bulk we propose the following ansatz \footnote{In Appendix \ref{Ap-0} we show an exact solution using separation of variables method.} (see Appendix \ref{Ap-1})

\begin{equation}
\label{ansatz}
\frac{\dot{a}}{n}=\lambda(t)a^{m/2},
\end{equation}
where substituting in the equations (\ref{friedmann3}) and (\ref{0-5}) immediately yields

\begin{eqnarray}
\label{ecuacionb}
b&=&\mathcal{C}a^{m/2}, \\ 
\label{completa}
a^{\prime}&=&\pm\ \mathcal{C}a^{m/2}\sqrt{\lambda^{2}a^{m}+k-\frac{\Lambda_{5}}{6}a^{2}+C_{_{DR}}a^{-2}},
\end{eqnarray}
where $\mathcal{C}$ is an integration constant and the term $\lambda^{2}$ behave as curvature, cosmological constant (CC) and dark radiation when $m=0,2,-2$ respectively.

On the other hand, cosmological observations (WMAP-7) confirms that the universe fits very well with the flat geometry $k\simeq0$. Equivalently, the additional degree of freedom corresponding to the dark radiation, can be constrained with the CMB observations to be no more than $\sim5\%$ of the universe density with the aim of allow nucleosynthesis \cite{Maartens:2010ar}.
Under the previous assertions, we elect that $k=C_{_{DR}}=0$ and the propositions that the five dimensional CC is negligible\footnote{See Appendix \ref{Ap-2} for a general solution in a not vacuum bulk.} 
 $\Lambda_{(5)}\simeq0$.

So, integrating \eqref{completa} and using the boundary conditions (\ref{salto1}-\ref{salto4}) with the addition of the equation \eqref{condicion1}, we obtain

\begin{eqnarray}
\label{solucion1}
a(t,y)&=&a_{0}\left[1+(m-1)\frac{\kappa^{2}_{(5)}}{6}\rho_{0}b_{0}y\right]^{1/(1-m)}, \\	
\label{solucion2}	
n(t,y)&=&n_{0}\left[1+\left(\frac{m}{2}+2+3\omega_{0}\right)\frac{\kappa^{2}_{(5)}}{6}\rho_{0}b_{0}y\right]\times\\
&&\left[1+(m-1)\frac{\kappa^{2}_{(5)}}{6}\rho_{0}b_{0}y\right]^{m/(2-2m)}, \nonumber\\
\label{solucion3}
b(t,y)&=&b_{0}\left[1+(m-1)\frac{\kappa^{2}_{(5)}}{6}\rho_{0}b_{0}y\right]^{m/(2-2m)}.
\end{eqnarray}
where $m\neq1$. Notice that for $m=0$ we recover the linear solutions founded by Langlois~\cite{Binetruy:1999ut} and for $m=2$ and $\omega_{0}=-1$, we find a conformal RS metric\footnote{Even though it is similar to the metric for a RS cosmology, we need non-constant values of $a_{0}$, $n_{0}$ and $b_{0}$. Note in this case, from Eq. \eqref{conexion0}, we also have $\rho_{0}=-\rho_{c}$. To recover the RS solutions, we should keep $\Lambda_{5}\neq 0$ and $\lambda=0$ in Eq. \eqref{completa}.}.  

 \section{Cosmological Analysis Between the Branes.}\label{Sec-3}
 
In this section, we study the cosmological results associated with the model using different components immersed in the branes. 
 
Starting with the solutions (\ref{solucion1}-\ref{solucion3}) in equations (\ref{salto1}-\ref{salto4}) we can prove that the following expressions for $\rho_{c}$ and $\omega_{c}$ can be written as

\begin{eqnarray}
\label{conexion0}
\rho_{c}&=&-\rho_{0}\left[1+(m-1)\frac{\kappa^{2}_{(5)}}{6}\rho_{0}b_{0}y_{c}\right]^{(m-2)/(2-2m)}\\
\label{conexion1}
\omega_{c}&=&\frac{\omega_{0}+(\frac{m}{2}+2+3\omega_{0})(\frac{m}{6}-1)\frac{\kappa^{2}_{(5)}}{6}\rho_{0}b_{0}y_{c}}{1+(\frac{m}{2}+2+3\omega_{0})\frac{\kappa^{2}_{(5)}}{6}\rho_{0}b_{0}y_{c}}
\end{eqnarray}
where we have supposed that the two branes are dominated by a perfect fluid component, \emph{i.e.} $p_{0}=\omega_{0}\rho_{0}$ and $p_{c}=\omega_{c}\rho_{c}$. 

Now, turning to Eq.~(\ref{friedmann3}), and considering the aforementioned particular case $k =\Lambda_{5} = C_{_{DR}}=0$, we find that

\begin{equation}
\label{Friedmann3a}
\frac{\dot{a}}{an}=\pm\frac{a'}{ab} \, .
\end{equation}
Taking into account the boundary conditions (\ref{salto1}) and (\ref{salto3}), we can write the Hubble parameter in the two branes as

\begin{eqnarray}
\label{friedmann3b}
\label{hubble0-0}
H_{0}&\equiv&\frac{\dot{a}_{0}}{a_{0}}=\mp\
\frac{\kappa^{2}_{(5)}}{6}n_{0}\rho_{0} \\ 
\label{hubbleC-0}
H_{c}&\equiv&\frac{\dot{a}_{c}}{a_{c}}=\pm\
\frac{\kappa^{2}_{(5)}}{6}n_{c}\rho_{c}
\end{eqnarray}
To be consistent with a FRW metric on the $y=y_{c}$-brane we impose $n_{c}=1$. So, using (\ref{solucion2}) and (\ref{conexion0}), we obtain

\begin{eqnarray}
\label{hubble0-1}
H_{0}&=&\mp\
\frac{\kappa^{2}_{(5)}}{6}\rho_{0}\frac{\left[1+(m-1)\frac{\kappa^{2}_{(5)}}{6}\rho_{0}b_{0}y_{c}\right]^{-m/(2-2m)}}{\left[1+\left(\frac{m}{2}+2+3\omega_{0}\right)\frac{\kappa^{2}_{(5)}}{6}\rho_{0}b_{0}y_{c}\right]} \\
\label{hubbleC-1}
H_{c}&=&\mp\
\frac{\kappa^{2}_{(5)}}{6}\rho_{0}\left[1+(m-1)\frac{\kappa^{2}_{(5)}}{6}\rho_{0}b_{0}y_{c}\right]^{(m-2)/(2-2m)}.
\end{eqnarray}
Eq.~(\ref{hubble0-0}), together with Eq.~(\ref{conservacion}), gives a solution for $\rho_{0}(t)$ when $\omega_{0}$ is a constant, and from it we obtain the evolution in time for $H_{c}(t)$. The sign $\mp$ is chosen such that we obtain an expanding universe within the $y=y_{c}$-brane. In the next, we consider two interesting cases when $m=0$\footnote{In Appendix \ref{Ap-3} we analyze the case $m\neq 0$ in two different regimes for $a_{0}$}.

\subsection{Single component on the y=0-brane with $\omega_{0}$ constant}

\begin{figure}[htp]
\centering
\includegraphics[scale=0.35]{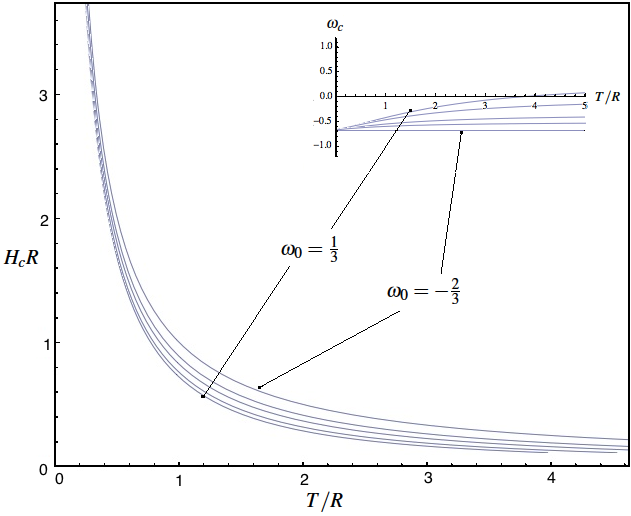} \includegraphics[scale=0.35]{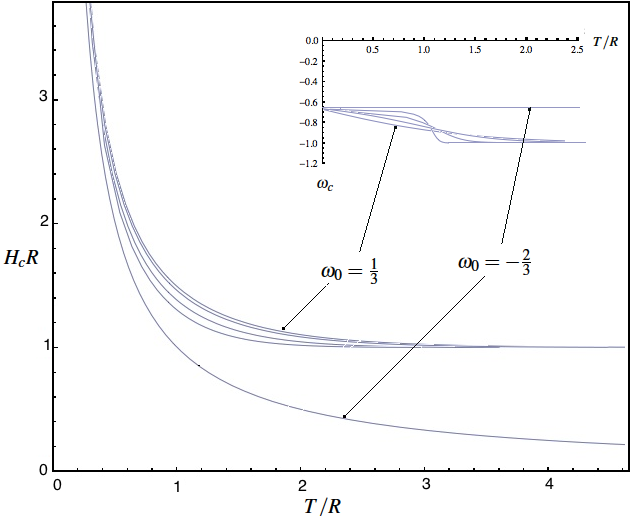}
\caption{Evolution of $H_{c}R$ as a function of $T/R$ when $m=0$ and $\omega_{0}\in[-2/3,1/3]$, see Eqs.~(\ref{casom=0}). (Upper) The case $0<X<1$ and sign $(-)$; note that $H_{c}R$ approaches zero when $T/R>>1$. (Bottom) The case $X>1$ and sign $(+)$; here $H_{c}R$ approaches the unity when $T/R>>1$, except  for $\omega_{0}=-2/3$, which is shown separately. The insets in both figures show, from Eq.~(\ref{conexion1}), the evolution of $\omega_{c}$ as a function of $T/R$ in each case.}
\label{fig:E+}
\end{figure}

As an example, let us consider the case $m=0$ and $\omega_{0}=const$ \emph{i. e.} a bulk metric that is linear in $y$, and the $y=0$-brane is dominated by a single component. From Eqs.~(\ref{conservacion}), (\ref{hubble0-0}) and (\ref{hubbleC-0}), we obtain
       
\begin{eqnarray}\label{casom=0}
\frac{T}{R}&=&\pm\frac{1-X^{-1}+(2+3\omega_{0})lnX}{3\ (1+\omega_{0})} \\
H_{c}R&=&\mp\ \frac{X}{1-X},
\end{eqnarray}
where $T=t-t_{*}$ , $X=\rho_{0}/\rho_{0*}$ , $\rho_{0*}=\rho_{0}(t_{*})$ , $R=b_{0}y_{c}$ is the radius of compactification, and $\omega_{0}\neq-1$. The constant $t_{*}$ is an epoch in which $\rho_{0*}^{-1}=\kappa^{2}_{(5)}b_{0}y_{c}/6$, \emph{i. e.} proportional to $R$. There exist two cases that guarantee positive solutions for $H_{c}$ in the $y=y_{c}$-brane, namely $0<X<1$ when the sign is $-1$ and $X>1$, when $+1$, which are shown in Fig. \ref{fig:E+}.

\subsection{Inflationary scalar field component on the y=0-brane}

\begin{figure}[h]
\centering
\includegraphics[width=0.45\textwidth]{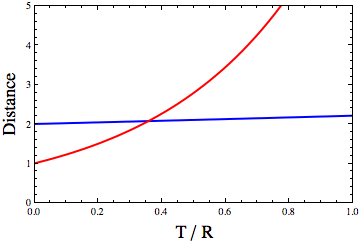} \includegraphics[width=0.45\textwidth]{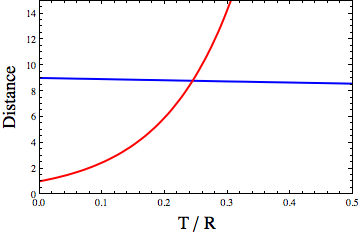} 
\caption{Evolution of $H_{c}R$ (Blue) and $a_{c}/a_{c*}$ (Red) as a function of $T/R$ (see Eqs. (\ref{scalar-solutions}) and (\ref{ln})). (Upper) the case when $\sigma=0.1$, $\gamma=2$ and the chosen sign is (+). (Bottom) the case when $\sigma=0.1$, $\gamma=0.9$ and the chosen sign is ($-$). Note in both cases, we have an exponential expanding scale factor, while the Hubble radius remains constant in the given interval.}
\label{Scalar}
\end{figure}

Consider now $m=0$ and a scalar field $\phi$ on the y=0-brane with evolution equations given by

\begin{eqnarray}\label{scalar-eqs} 
&&\rho_{0}=\frac{\dot{\phi}^{2}}{2}+V(\phi), \\
&&p_{0}=\frac{\dot{\phi}^{2}}{2}-V(\phi), \\
&&\ddot{\phi}+3H_{0}\dot{\phi}+\partial_{\phi}V(\phi)=0,
\end{eqnarray}
with a power-law inflation described by the exponential potential, $V(\phi)=V_{0}e^{B\phi}$. In the slow-roll limit, $\dot{\phi}^{2}\ll2V(\phi)$ and $\ddot{\phi}\ll3H_{0}\dot{\phi}$, the Eqs. (\ref{hubble0-1}) and (\ref{hubbleC-1}) are similar; so, integrating them, it is straightforward to obtain

\begin{eqnarray}\label{scalar-solutions}
H_{c}R&=&\frac{\pm\gamma}{\gamma-1}e^{\pm\sigma \frac{T}{R}}, \\ \label{ln}
ln\left(\frac{a_{c}}{a_{c*}}\right)&=&\frac{\gamma}{\sigma(\gamma-1)}\left[e^{\pm\sigma\frac{T}{R}}-1\right],
\end{eqnarray}
where

\begin{equation}
T=t-t^{*},\ \ \gamma=\frac{1}{6}\kappa^{2}_{(5)}RV_{0}e^{B\phi_{*}}, \ \ \sigma=\frac{2R}{\kappa^{2}_{(5)}}B^{2}.
\end{equation}
Note, when the sign is ($+$), $\sigma T/R\ll1$ and $\gamma>1$ we have an expanding universe in our brane, being $\phi_{*}$ constant in a $t^{*}$ time. Another possibility to have an expanding brane is to take sign ($-$), $0<\sigma<1$ and $0<\gamma<1$.  As an example, It is to showed in Fig. \ref{Scalar} the evolution of the Hubble parameter and the scale factor in the visible brane, firstly for $\sigma=0.1$, $\gamma=2$ and sign ($+$), and later for $\sigma=0.1$, $\gamma=0.9$ and sign ($-$). 

Analogously, a useful expression to understand the inflationary behavior in the brane world context it is through the $\epsilon$ and $\eta$ parameters written in the following form \cite{Maartens:2010ar}

\begin{equation}\label{epsilon}
\epsilon=\left|\frac{2}{\kappa^{4}_{(5)}}\frac{V'(\phi)^{2}}{V(\phi)^{3}}\right|=\left|\frac{\sigma}{6}\left( 1-\frac{\gamma-1}{\gamma}e^{\mp\sigma T/R}\right)\right|\ll1
\end{equation}

\begin{equation}\label{eta}
\eta=\left|\frac{-4}{\kappa^{4}_{(5)}}\frac{V''(\phi)}{V(\phi)^{2}}\right|=2\epsilon\ll1
\end{equation}

\begin{figure}[h]
\centering
\includegraphics[width=0.45\textwidth]{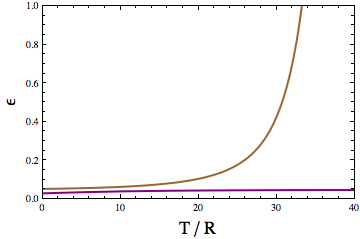}
\caption{Evolution of $\epsilon$ as a function of $T/R$ (see Eq. (\ref{epsilon})). (Purple) The case when $\sigma=0.1$, $\gamma=2$ and the chosen sign is ($-$); in this case the brane will expand forever. (Brown) the case when $\sigma=0.1$, $\gamma=0.9$ and the chosen sign is ($+$); in this case the inflation will end.}
\label{fig-eta}
\end{figure}

Immediately, we observe, from Fig. \ref{fig-eta} that the election of the sign $\pm$ is crucial to obtain an inflationary or not inflationary brane. In the case (+), we note, from Eq. (\ref{epsilon}), the brane expands forever, while in the case ($-$), we can note that the expansion will eventually end.

Finally, the numbers of e-foldings in the model can be computed using the following established equation 

\begin{equation}
N=-3\int\frac{H_{c}^{2}}{\partial_{\phi}V(\phi)}d\phi\simeq-\frac{3}{V_{0}B}\int H_{c}^{2}e^{-B\phi}d\phi,
\end{equation}
where $H_{c}$ can be obtained form the equation \eqref{scalar-solutions} assuming the inflationary exponential potential.

Note, the exponential expansion behavior is recovered in our brane, as a result of having a scalar field in the hidden brane. If we suppose in our brane there exist only Standard model fields, they are not playing a crucial roll at least until inflation ends. In the next section we explore about the possibility to have dark matter fields in the hidden brane, such that they induce the behavior of dark matter in our brane.

\section{Conclusions and Remarks}\label{concl}
We had to showed there exist a relationship between the cosmologies in both branes, strictly speaking the energy density behavior in one brane affects the energy in the another one such as in the Randall Sundrum cosmology which is a particular case in our analysis; we have recovered, and generalized, the expression found by Langlois \cite{Binetruy:1999ut} for two branes and the evolution for the Hubble parameter in our brane is recovered as a consequence of the 5-D geometry (given by $m$) and the dominant component in the hidden brane (given by $\omega_{0}$). We focusses only in the vacuum bulk case because it has an exact solution. Supposing the hidden brane is dominated by a single component, ($\omega_{0}=0$) and a particular geometry ($m=0$), we obtain exact solutions for the Hubble parameter as a function in time. The scalar field component in the hidden brane is also considered, where the potential is exponential in $\phi$, and, as consequence, an exponential expansion is recovered in our brane. The behavior of a scalar field dark matter in the hidden brane and the visible effects in our brane will be studied in a future works.  

\begin{acknowledgments}
We want to thank M. A. Garc\'ia-Aspeitia for useful comments. We acknowledge partial support by SNI-M\'exico, CONACyT research grant J1-60621-I, COFAA-IPN and SIP-IPN grant 20121648. This work was partially supported by PIFI, PROMEP, DAIP-UG, CONACyT M\'exico under grant 167335, and the Instituto Avanzado de Cosmolog\'ia (IAC) collaboration.
\end{acknowledgments} 

\appendix

\section{5-D Einstein equations} \label{Ap-1}

The five-dimensional non zero Einstein tensors, $\tilde{G}_{_{AB}}$, for the metric (\ref{metrica}) are written as 
\begin{eqnarray}
\tilde{G}_{00}&=&3\frac{\dot{a}}{a}\left(\frac{\dot{a}}{a}+\frac{\dot{b}}{b}\right)-3\frac{n^{2}}{b^{2}}\left[\frac{a''}{a}+\frac{a'}{a}\left(\frac{a'}{a}-\frac{b'}{b}\right)\right]+3k\frac{n^2}{a^2}, \nonumber\\ 
\label{einstein1}
\tilde{G}_{ij}&=&\frac{a^{2}}{b^{2}}\delta_{ij}\left\{\frac{a'}{a}\left(\frac{a'}{a}+2\frac{n'}{n}\right)-\frac{b'}{b}\left(\frac{n'}{n}+2\frac{a'}{a}\right)\right\}\nonumber\\
&+&\frac{a^{2}}{b^{2}}\delta_{ij}\left\{2\frac{a''}{a}+\frac{n''}{n}\right\}+\frac{a^{2}}{n^{2}}\delta_{ij}\left\{\frac{\dot{a}}{a}\left(-\frac{\dot{a}}{a}+2\frac{\dot{n}}{n}\right)\right\}\nonumber\\
&+&\frac{a^{2}}{n^{2}}\delta_{ij}\left\{-2\frac{\ddot{a}}{a}+\frac{\dot{b}}{b}\left(-2\frac{\dot{a}}{a}+\frac{\dot{n}}{n}\right)-\frac{\ddot{b}}{b}\right\}-k\delta_{ij}, \nonumber\\
\label{einstein2}
\tilde{G}_{05}&=&3\left(\frac{\dot{a}}{a}\frac{n'}{n}+\frac{\dot{b}}{b}\frac{a'}{a}-\frac{\dot{a}'}{a}\right), \nonumber\\
\label{einstein3}
\tilde{G}_{55}&=&3\frac{a'}{a}\left(\frac{a'}{a}+\frac{n'}{n}\right)-3\frac{b^{2}}{n^{2}}\left[\frac{\ddot{a}}{a}+\frac{\dot{a}}{a}\left(\frac{\dot{a}}{a}-\frac{\dot{n}}{n}\right)\right]-3k\frac{b^2}{a^2}\nonumber.
\end{eqnarray}
In the interval $0<y<y_{c}$, the Einstein equations, $\tilde{G}_{AB}=\kappa^{2}_{(5)}\tilde{T}_{AB}$, reads

\begin{eqnarray}\label{einstein-compactas}
\label{e-c-0}
\frac{Y'}{b}-\frac{XW}{ab}&=&k-\frac{1}{3}a^{2}\Lambda_{5},\\
\label{e-c-1}
\frac{Y'}{b}-\frac{\dot{X}}{n}+\frac{a}{bn}(Z'-\dot{W})&=&k-a^{2}\Lambda_{5},\\
\label{e-c-2}
W&=&a\frac{X'}{Y},\\ 
\label{e-c-3}
Z&=&a\frac{\dot{Y}}{X},\\
\label{e-c-4}
\frac{\dot{X}}{n}-\frac{YZ}{an}&=&-k+\frac{1}{3}a^{2}\Lambda_{5},
\end{eqnarray}
where 
\begin{equation}\label{xyzw}
X=\frac{a\dot{a}}{n}, \ \ Y=\frac{aa'}{b}, \ \ Z=\frac{(an)'}{b}, \ \ W=\frac{\dot{(ab)}}{n}.
\end{equation}
Substituting (\ref{e-c-2}) in (\ref{e-c-0}) and integrating, we obtain 
\begin{equation}\label{fr-3}
Y^{2}-X^{2}=a^{2}k-a^{4}\frac{\Lambda_{5}}{6}+\emph{constant},
\end{equation}
which is Eq. (\ref{friedmann3}) if we identify the constant with $C_{_{DR}}$. Note that the Eq. (\ref{e-c-2}) and Eq. (\ref{e-c-3}) are both the same than Eq. (\ref{0-5}). 

An exact solution for the metric coefficients is obtained when $X=\lambda(t)f(a)$. Substituting this ansatz in (\ref{xyzw}), this implies, 
 \begin{equation}\label{sol-n}
n=\frac{a\dot{a}}{\lambda(t)f(a)}.
\end{equation}  
Now, we can integrate Eq. (\ref{0-5}) to give,
\begin{equation}\label{sol-b}
b=\mathcal{C}\frac{f(a)}{a},
\end{equation}  
Finally, from (\ref{fr-3}) we obtain a differential equation for $a$

\begin{equation}\label{sol-a}
a^{\prime}=\pm\ \mathcal{C}\frac{f(a)}{a}\sqrt{\lambda(t)^{2}\left(\frac{f(a)}{a}\right)^{2}+k-\frac{\Lambda_{5}}{6}a^{2}+\frac{C_{_{DR}}}{a^{2}}}.
\end{equation} 
Usually, to have a stabilized bulk, $b$ is taken to be the unity, and therefore, $f(a)/a=\mathcal{C}^{-1}$  \cite{Langlois,Mukohyama:1999qx}. In this paper, we take a general form $f(a)/a=a^{m/2}$, which is the case of non-static internal dimensions.

\section{General solutions with separation of variables}\label{Ap-0}
We can solve (\ref{friedmann3}) and (\ref{0-5}) exactly when the metric coefficients can be written as
\begin{eqnarray}\label{separacion1}
a&=&a_{t}(t)a_{y}(y)\\
n&=&n(y)\\
b&=&b(t)
\end{eqnarray}
Under this ansatz and using (\ref{salto1}) and (\ref{salto4}), Eq. (\ref{0-5}) gives the solutions
\begin{eqnarray}\label{solsep1}
n&=&k_{1}a_{y}^{\alpha}\\
b&=&k_{2}a_{t}^{1-\alpha}\\
\alpha&=&-(2+3\omega_{0})=-(2+3\omega_{c})
\end{eqnarray}
where $k_{1},\ k_{2},\ \alpha$ are all constants. Note in this case, $\omega_{0}=\omega_{c}$ both constants, and therefore, the evolution is similar in both branes. Using this results in (\ref{friedmann3}), the metric coefficient $a(t,y)$ must satisfy the follow differential equation.
\begin{eqnarray}\label{separacion2}
&&\left(\frac{a_{y}'}{k_{2}a_{y}}\right)^2a^{2(\alpha-1)}_{t}-\left(\frac{\dot{a}_{t}}{k_{1}a_{t}}\right)^2a^{-2\alpha}_{y}=\nonumber\\  
&=&ka_{t}^{-2}a_{y}^{-2}-\frac{\Lambda_{5}}{6}+C_{DR}a_{t}^{-4}a_{y}^{-4}, 
\end{eqnarray}
This last equation is solvable only in four cases: 
\begin{eqnarray}
1.\ &k&\neq0;\ \ \Lambda_{5}=C_{DR}=0;\ \ \alpha=0,1\nonumber \\  
2.\ &\Lambda_{5}&\neq0;\ \ K=C_{DR}=0;\ \ \alpha=0,1\nonumber \\ 
3.\ &C_{DR}&\neq0;\ \ K=\Lambda_{5}=0;\ \ \alpha=3,-3\nonumber \\
4.\ &k&=\Lambda_{5}=C_{DR}=0;\ \ \alpha=any\nonumber
\end{eqnarray}
The signs for the constants is determinant for the solution.  

\section{General Solutions in a not vacuum bulk}\label{Ap-2}
 
Returning to Eq. (\ref{completa}), the general metric that solve the 5-D Enistein equations must satisfy
\begin{equation}\label{integral}
\int_{a_{0}}^{a}\frac{a^{1-m/2}}{\sqrt{\lambda^{2}a^{m+2}-\frac{\Lambda_{5}}{6}a^{4}+ka^{2}+C_{_{DR}}}}da=\pm\mathcal{C}y.
\end{equation}
This integrals is only solvable for $m=0, 2, -2$. The case $m=0$ is easy to solve and it is showed in \cite{Langlois} the general solution. The other two cases can be solving only with elliptic integrals. 

For $m=+2$,
\begin{equation}\label{elliptic1}
F\left[sin^{-1}\hat{a},\sqrt{R_{-}/R_{+}}\right]=\pm\sqrt{R_{+}(\lambda^{2}-\Lambda_{5}/6)}\mathcal{C}y,
 \end{equation}
where

\begin{eqnarray}
\hat{a}&=&\frac{a}{\sqrt{R_{-}}},\nonumber \\
R_{\pm}&=&\frac{-k\pm \sqrt{k^{2}-4C_{_{DR}}(\lambda^{2}-\Lambda_{5}/6)}}{2(\lambda^{2}-\Lambda_{5}/6)},\nonumber \\
0&<&\frac{R_{-}}{R_{+}}<1,\nonumber \\
0&<&\lambda^{2}-\frac{1}{6}\Lambda_{5}, \nonumber 
\end{eqnarray}

For $m=-2$
\begin{eqnarray}\label{elliptic2}
&&F\left[sin^{-1}\hat{a},\sqrt{R_{-}/R_{+}}\right]-E\left[sin^{-1}\hat{a},\sqrt{R_{-}/R_{+}}\right]\nonumber\\ 
&&=\pm\sqrt{\frac{-\Lambda_{5}}{6R_{+}}}\mathcal{C}y, 
 \end{eqnarray}
where
\begin{eqnarray}
R_{\pm}&=&\frac{-k\pm \sqrt{k^{2}+4\Lambda_{5}(C_{_{DR}}+\lambda^{2})/6}}{-2\Lambda_{5}/6},\nonumber \\
0&<&\frac{R_{-}}{R_{+}}<1,\nonumber \\
0&<&-\frac{1}{6}\Lambda_{5}, \nonumber
\end{eqnarray}
being $F(x,y)$ and $E(x,y)$ the incomplete elliptic integrals of the first and second kind respectively. 

\section{Two limit cases}\label{Ap-3}

Solving Eq.~(\ref{conservacion}) for  $\omega_{0}$ constant, and using (\ref{ecuacionb}), we can write 
\begin{equation}\label{conservacionC}
\rho_{0}b_{0}=\Gamma C a_{0}^{m/2-3(1+\omega_{0})}.
\end{equation}
We want to know how is the evolution for the Hubble parameter $H_{c}$ as a function of $a_{c}$, when $a_{0}<<1$ and later when $a_{0}>>1$. Note, the sign of $m/2-3(1+\omega_{0})$ determines the convergence or divergence of (\ref{conservacionC})

\subsection{$a_{0}<<1$ case}
We are interested in a \emph{very small} value for $a_{0}$ such that 
\begin{equation}\label{l知itealta}
\left|(m-1)\frac{\kappa^{2}_{(5)}}{6}\rho_{0}b_{0}y_{c}\right|>>1  
\end{equation}
when $m/2-3(1+\omega_{0})<0$, and 
\begin{equation}\label{l知itebaja}
\left|(m-1)\frac{\kappa^{2}_{(5)}}{6}\rho_{0}b_{0}y_{c}\right|<<1  
\end{equation}
when $m/2-3(1+\omega_{0})>0$. Using this limits in  (\ref{solucion1}) and (\ref{hubbleC-1}) with aid of (\ref{conservacionC}), we obtain,
\begin{equation}
 H_{c} =\frac{\dot{a_{c}}}{a_{c}}= \left \{ \begin{matrix} 
-\epsilon\frac{1}{(m-1)Cy_{c}}a_{c}^{-m/2} &;\ \omega_{0}>\frac{m}{6}-1\\ 
 \\
-\epsilon\ \frac{\kappa^{2}_{(5)}}{6}\Gamma a_{c}^{-3(1+\omega_{0})} &;\ \omega_{0}<\frac{m}{6}-1
\end{matrix}\right.  
\end{equation}

\subsection{$a_{0}>>1$ case}
In this case, we are interested in a \emph{very large} value for $a_{0}$ such that 
\begin{equation}\label{l知itealta}
\left|(m-1)\frac{\kappa^{2}_{(5)}}{6}\rho_{0}b_{0}y_{c}\right|>>1  
\end{equation}
when $m/2-3(1+\omega_{0})>0$, and 
\begin{equation}\label{l知itebaja}
\left|(m-1)\frac{\kappa^{2}_{(5)}}{6}\rho_{0}b_{0}y_{c}\right|<<1  
\end{equation}
when $m/2-3(1+\omega_{0})<0$. Using this limits in  (\ref{solucion1}) and (\ref{hubbleC-1}) with aid of (\ref{conservacionC}), we obtain,
\begin{equation}
 H_{c} =\frac{\dot{a_{c}}}{a_{c}}= \left \{ \begin{matrix} 
-\epsilon\frac{1}{(m-1)Cy_{c}}a_{c}^{-m/2} &;\ \omega_{0}<\frac{m}{6}-1\\ 
 \\
-\epsilon\ \frac{\kappa^{2}_{(5)}}{6}\Gamma a_{c}^{-3(1+\omega_{0})} &;\ \omega_{0}>\frac{m}{6}-1
\end{matrix}\right.  
\end{equation}

This last result shows the inherited cosmology in the brane $y=y_{c}$ when the bulk have a particular topology (given by $m$) and when the brane $y=0$ is dominated by one only component (given by $\omega_{0}$) in a vacuum bulk.

\bibliographystyle{apsrev} 
\bibliography{librero1}

\end{document}